\documentclass[12pt]{article}

\usepackage{amsmath,amsfonts,epsfig,color,latexsym}

\newcommand{\be}{\begin{equation}}
\newcommand{\ee}{\end{equation}}
\newcommand{\bea}{\begin{eqnarray}}
\newcommand{\eea}{\end{eqnarray}}

\renewcommand{\theequation}{\thesection.\arabic{equation}}

\let\newsection=\section
\renewcommand{\section}{\setcounter{equation}{0}\newsection}

\begin{document}

\begin{flushright}
hep-th/0601182\\
BROWN-HET-1461
\end{flushright}
\vskip.5in

\begin{center}

{\LARGE\bf Constant H field, cosmology and faster than light solitons}
\vskip 1in
\centerline{\Large Horatiu Nastase}
\vskip .5in

\end{center}
\centerline{\large Brown University}
\centerline{\large Providence, RI, 02912, USA}

\vskip 1in

\begin{abstract}

{\large We analyze the possibility of having a constant spatial NS-NS 
field, $H_{123}$. Cosmologically, it will act as stiff matter,
and there will be very tight constraints on the possible 
value of $H_{123}$ today. However, it will give a noncommutative structure
with an {\em associative} star product of the type $\theta^{ij}=\alpha 
\epsilon^{ijk} x^k$. This will be a fuzzy space with constant radius 
slices being fuzzy spheres. We find that gauge theory on such a space 
admits a noncommutative soliton with galilean dispersion relation, thus 
having speeds arbitrarily higher than c. This is the analogue of the 
Hashimoto-Itzhaki construction at constant $\theta$, except that one has 
fluxless solutions of arbitrary mass. A holographic description 
supports this finding. We speculate thus that the presence of constant 
(yet very small) $H_{123}$, even though otherwise virtually 
undetectable could still imply the existence of faster than 
light solitons of arbitrary mass
(although possibly quantum-mechanically unstable). The spontaneous
Lorentz violation given by $H_{123}$ is exactly the same one already 
implied by the FRW metric ansatz.
}

\end{abstract}

\newpage

\section{Introduction}

A constant NS-NS B field has zero field strength, thus could maybe have been 
considered just a choice of gauge, except it was shown in detail in 
\cite{sw} that in a certain limit it makes the space noncommutative. 
Noncommutative geometry would imply certain spontaneous violation of 
Lorentz invariance, given that, for instance, $\theta^{12}$ is a constant 
tensor (making the direction 3 preferred). As such, there were limits put 
on the size of $\theta$. A constraint based on a noncommutative version 
of QCD drives $\theta$ down to about $(10^{14}GeV)^{-2}$ \cite{mpr}, whereas a 
more conservative constraint based on just noncommutative QED gives it 
the maximal possible value of $(10\; TeV)^{-2}$ \cite{chklo} (see also
\cite{cst,cpst} for related bounds). 
More interestingly, noncommutativity
also predicts the existence of gauge theory solitons 
\cite{gms,poly,bak,agms,hkl,gn}, which were shown in 
a very interesting paper \cite{hi} to have 
galilean dispersion relations, thus moving at speeds arbitrarily higher than 
the speed of light, but only in the noncommutative directions
(in fact, it was even shown in \cite{llt}, that low momentum 
wavepackets can move faster than c due to UV/IR mixing). 
It is not clear what the phenomenological implications of this fact are, 
yet there is something disconcerting about such a strong breaking of 
rotational invariance (clearly choosing, be it even locally, a preferred 
direction in space). It certainly seems to imply that the B field cannot 
be constant over the whole Universe. 

One can however also ask what happens if one chooses a constant H=dB field. 
This is a solution of the equations of motion in Freedman-Robertson-Walker
(FRW) geometry, as we will 
see, but also has a more satisfying feature compared to a constant B field:
one can choose a constant field with only 3+1d spatial indices, i.e. $H_{123}$,
which doesn't break any more Lorentz invariance than the one already 
assumed in the FRW metric ansatz, i.e. a choice of time slicing, physically 
corresponding to the cosmic microwave background radiation (CMBR) reference
frame. Indeed, there are only two types of Lorentz violations that are 
consistent with the FRW ansatz: choosing a nonzero constant antisymmetric 
tensor $T_{123}$, or its 4d dual, a nonzero $V_0$ component of a vector.
One can take two points of view about that: either to say that $T_{123}$ 
and $V_0$ are consistent with the FRW ansatz, or maybe even to take it, 
in the case of $T_{123}$, as a seed for the FRW cosmology, thus as a 
possible explanation for our expanding 3+1 dimensional space. We will 
not explore this possibility here, but it is conceivable that if one 
has a constant tensor $T_{123}$ at the Big Bang, it could provide the 
initial condition for our FRW space.
Among the fields of string theory, the NS-NS fields are universal, so 
they are model independent, and the only field that satisfies the 
previous requirement is $H_{\mu\nu\rho}$.
In this paper we will try to explore the consequences of a constant $H_{123}$.

Constant H field will imply a varying B field, and one can ask whether 
one has also a noncommutative structure. A space-time varying noncommutativity
was analyzed before, first for a time-dependent one in \cite{hs}, shown to 
be consistent in \cite{dn}, then for a space-dependent one in \cite{lnr}, 
and further studied in \cite{rs,ht} (see also \cite{clo}). 
We will use cosmology to impose 
constraints on the constant $H_{123}$ and find that it is small enough that 
the noncommutative structure will be of the type $\theta^{ij}=\alpha \epsilon
^{ijk}x^k$. The faster than light solitons on the fuzzy sphere were treated
in \cite{hk}, but that case, even if similar to ours, is different in that
it deals still with 2d solitons, whereas we deal with 3d solitons. It would 
still be interesting to see if there are any connections though. 
After this paper was finished, we became aware of \cite{hls}, where the 
same fuzzy space we considered was analyzed, and called $R^3_{\lambda}$.

We will find that $H_{123}$ behaves like stiff matter of negative energy 
density, and thus falls off on cosmological time scales as $1/a^6$, thus it 
cannot have much impact on recent cosmology, but at most on the initial 
conditions. Yet given the example of \cite{hi}, there could be still 
measurable phenomenological implications in the form of faster than light 
solitons. We will look for such solitons in the gauge theory on the 
space with $\theta^{ij}=\alpha \epsilon^{ijk}x^k$ and see if we get the 
same result from a holographic description. 

The paper is organized as follows. In section 2 we analyze the possible 
noncommutative structure, using cosmology to set bounds on H and see 
the implications. In section 3 we ask whether $H_{123}$ can have any 
important effects on cosmology. In section 4 we build noncommutative 
solitons in the gauge theory on the $\theta^{ij}=\alpha \epsilon^{ijk}x^k$
space, paralleling the analysis in \cite{hi}. In section 5 we derive a 
holographic description of the gauge theory and check that our solitons 
can be described in it. In section 6 we conclude. The Appendix contains 
a review of the analysis in \cite{hi}, for use in the paper.

\section{Constant H field and FRW cosmology}

In FRW cosmology, one always breaks spontaneously Lorentz invariance, i.e. 
there is a preferred cosmological time, thus a preferred cosmological frame
(the rest frame of the CMBR). Thus one can obtain some superluminal motion
just from the cosmological evolution (see for instance \cite{mv}). 
Of course, rotational and translational 
invariance is preserved (the FRW solution is in fact the unique homogeneous 
and isotropic cosmology, i.e. the unique cosmological 
solution that preserves rotational and translational invariance). 
But that means that in FRW cosmology one can have a constant field 
$V_0$ (zero component of a vector) or $T_{123}$ (123 component of an 
antisymmetric $T_{(3)}$ field) without further spontaneous breaking of 
Lorentz invariance (i.e. by the vacuum solution). But we know we generically 
have such a field.

{\bf Energy-momentum tensor}

In string theory we always have a dilaton and a NS-NS B field, together with 
gravity, as the general bosonic closed string sector (NS-NS). 
It has the action 
\be
-\frac{1}{2k^2}\int \sqrt{g} [-R + \frac{(\partial \phi)^2}{2}+e^{-\phi}
\frac{H_{MNP}^2}{2}]
\ee
which
has as B equation of motion $d *H=0$, and $dH=0$ as Bianchi identity,
which means that H=constant (and $\phi$=constant) is a 
solution. In particular, we will be interested in the case 
$H_{123}$=constant. Then the H field energy momentum tensor will be 
(we put $\phi=0$)
\be
T_{\mu\nu} (H) =\frac{1}{2} g_{\mu\nu} H^2 +3 H_{\mu NP}{H_{\nu}}^{NP}
\ee
and with $g_{\mu\nu}=g_{ii} \delta _{ij}$ ($-g_{00}=g_{ii}=+1$) we have 
\be
T_{\mu\nu} (H) = \frac{g_{ii}}{2}diag (\Lambda; \Lambda; \Lambda, \Lambda)=
diag (\rho; p;p;p)
\ee
where $\Lambda\equiv H^2 = 6(H_{123})^2 (g^{ii})^3>0$. If we consider instead 
the (4 dimensional) dual field ($\tilde{H}=*_4H$) to be the relevant 
dynamical field, and
$\tilde{H}_0$=constant the relevant solution, with $\Lambda = H^2 = g^{00} 
(\tilde{H}_0)^2 <0$, then from 
\be
T_{\mu\nu} =-\frac{1}{2} g_{\mu\nu} H^2 +H_{\mu}H_{\nu}
\ee
we get now 
\be
T_{\mu\nu}= -\frac{g_{ii}}{2}diag (\Lambda; \Lambda; \Lambda, \Lambda)
\ee
except now $\Lambda <0$. Thus in both cases $T_{\mu\nu} = g_{ii}/2
diag(|\Lambda|, |\Lambda|, |\Lambda|, |\Lambda|)$, giving positive 
energy density and pressure in the form of stiff matter ($p=\rho$). 
We can try to construct a $T_{\mu\nu}$ 
with only energy density or only pressure. 

If we add a negative cosmological constant $-|\Lambda|/2$, i.e. 
\be
T_{\mu\nu} (\Lambda)= g_{ii}/2
diag(-|\Lambda|, |\Lambda|, |\Lambda|, |\Lambda|)
\ee
to the constant H field energy-momentum tensor, 
we get no energy and only   pressure, 
\be
T_{\mu\nu} = g_{ii} diag (0,  |\Lambda|,  |\Lambda|,  |\Lambda|)
\ee
while if we add a positive cosmological constant 
$T_{\mu\nu} (\Lambda) = g_{ii}/2
diag(|\Lambda|, -|\Lambda|, -|\Lambda|, -|\Lambda|)$
we get energy density with no pressure (dust matter):
\be
T_{\mu\nu} = g_{ii} diag (|\Lambda|, 0, 0, 0)
\ee

However, in both cases the equality cannot persist in time, as a cosmological
constant stays constant in time, whereas we will see that $T_{\mu\nu}(H)$
drops quickly with time. 

Before we go on, let us pause and try to understand the constant H-field 
in string theory. At this point, $H_{123}$ is just an arbitrary 
constant, and we didn't make any assumptions about how we will get this. 
We took a model independent approach, and assumed there will be such an 
H-field, as this was allowed by the classical supergravity equations of 
motion, and one can understand the $H_{123}$ as a limit of an H field on 
a spatial 3-sphere, $H_{ijk}=\epsilon_{ijk}, i, j, k\in S^3$, when the 
radius of the sphere becomes infinite. As such, this could provide the 
initial conditions for our FRW cosmology, by having  3 spatial dimensions, 
curled up in a sphere, expand due to the H flux. However, such a model 
would depend on the existence of a consistent string theory model at 
the Planck era, and there are possible problems with that\footnote{I 
thank Aki Hashimoto for pointing these out to me}. We will deal with 
open strings in the following, which assumes the existence of Dp branes 
with $9\geq p\geq 3$, such that our 3+1 dimensional world is embedded in them. 
However, for such branes (in particular, for D3 branes) in the presence of 
an H flux, there exists a baryon anomaly, which requires the addition of 
stretched branes a la Hanany-Witten. So a particular consistent picture 
involving our constant H field and Dp branes (such that our 3+1 dimensions 
are embedded in them) will be difficult to find.
However, we will go on in this model independent way, assuming that 
at least nowadays this will not be an important problem, since as we will 
obtain, $H_{123}$ needs to be extremely small.

{\bf Open string variables and noncommutative geometry}

Let us now see what kind of cosmology will open strings (on an assumed 
Dp brane that contains our 3+1 dimensions) feel in the 
H field backgroud. 
Consider a constant H field with spatial components only, i.e. 
 $H_{123}$= constant, living in flat Minkowski space. Then  we can find a 
gauge where (rescaling constants to 1) $B_{12}=x_3, B_{23}= x_1, B_{31}=x_2,
g_{ij}=\delta_{ij}$. 

One can define  open string variables $G$ and $\theta$ in the usual way 
(following Seiberg and Witten \cite{sw}, 
these will be the variables felt by open 
strings moving in the above background)  
by $(G+\theta/(2\pi \alpha '))^{ij}=(1/(g+2\pi \alpha 'B))^{ij}$
(but we put for the moment $2\pi \alpha '=1$), giving 
\bea
&&G^{ij}= \frac{1}{1+x_1^2+x_2^2 +x_3^2}\begin{pmatrix} 1+x_1^2 &
x_1 x_2 & x_1 x_3\\ x_1 x_2 & 1+ x_2^2 & x_2 x_3 \\ x_1 x_3 & x_2 x_3 & 1+ x_3 
^2 \end{pmatrix} \nonumber\\
&&\theta^{ij}=  \frac{1}{1+x_1^2+x_2^2 +x_3^2}\begin{pmatrix}
0 &- x_3 & x_2 \\ x_3 & 0 & - x_1\\ -x_2 & x_1 & 0 \end{pmatrix}
\label{openvar}
\eea
Inverting $G^{ij}$ to $G_{ij}$ one finds that the open string variables 
are (the closed string variables were $ ds^2 =d\vec{r}^2, H_{123}=3$)
\bea
&& ds^2 = (1+\vec{r}^2)d\vec{r}^2 - (\vec{r}\cdot d\vec{r})^2=
(1+\vec{r}^2)d\vec{r}^2- r^2 dr^2 \nonumber \\
&& \theta^{ij}= -\frac{1}{1+\vec{r}^2}\epsilon^{ijk} x^k
\label{openv}
\eea

Of course, as shown in \cite{sw}, these variables are the ones appearing 
in the open string n-point functions, but in order to have just a 
noncommutative geometry structure, one needs to decouple the $\alpha '$ 
corrections from the effective action. This was done in \cite{sw} (for the 
constant B field case) by taking the limit $\alpha ' \sim \sqrt{\epsilon}
\rightarrow 0$, $ g_{ij}\sim \epsilon\rightarrow 0$ (closed string metric 
in the noncommutative directions), and keeping $G$ (open string metric) 
and $\theta, B$ fixed. Then one obtains, for instance that in the limit, 
$\theta^{ij}=(B^{-1})^{ij}$ (matrix inverse).  

But first of all, notice that a 3 dimensional antisymmetric matrix has no 
inverse, as one can explicitly check (unlike, say, a 2 dimensional or 
4 dimensional antisymmetric matrix). So clearly something special happens
when 3 coordinates are involved. We can gain a better understanding of this 
fact by looking at the case studied in \cite{lnr}, where only $B_{23} $ 
was nonzero. Specifically, there one had the metric and B field 
\bea
&& ds^2 = d\tilde{y}_1^2 +\frac{d\tilde{y}_2^2 +d\tilde{y}_3^2}{
1+(\tilde{\alpha}\tilde{y}_1/l_s^2)^2}\nonumber\\
&& B= -\frac{\tilde{\alpha}\tilde{y}_1/l_s^2}{
1+(\tilde{\alpha}\tilde{y}_1/l_s^2)^2}\frac{d\tilde{y}_2\wedge d\tilde
{y}_3}{l_s^2}
\eea
and after going to open string variables by $(G+\theta/l_s^2)^{ij}= 1/(
g+l_s^2B)_{ij}$ one obtains 
\be
ds^2= d\tilde{y}_1^2+d\tilde{y}_2^2 +d\tilde{y}_3^2;\;\;\;\;
\theta^{23}=\tilde{\alpha}\tilde{y}_1
\ee
One can check that the $l_s\rightarrow 0$ limit is exactly the Seiberg-Witten
limit  for decoupling of $\alpha '$ corrections (the closed string metric 
in the 2,3 directions goes like $l_s^4$, whereas B, G, $\theta$ are fixed),
however the open string variables are actually {\em independent of $l_s$}!
So we don't {\em need} to take $l_s\rightarrow 0$ to obtain the correct 
G and $\theta$!

Moreover, we immediately recognize that the rotationally invariant form of 
this case is just the case we were analyzing in (\ref{openvar}), except 
taken in reverse (exchange $(G,\theta)$ for $(g,B)$). Specifically, 
the closed string metric and B field
\bea
&& ds^2= \frac{1}{1+\alpha^2 \vec{r}^2/l_s^4}(d\vec{r}^2+\alpha^2(\vec{r}
\cdot d\vec{r})^2/l_s^4)= dr^2 +\frac{r^2}{1+\alpha^2 r^2 /l_s^4}d\Omega_2^2
\nonumber\\
&& B=-\frac{\alpha /l_s^2}{1+\alpha^2 \vec{r}^2/l_s^4}\epsilon^{ijk} x^i
\frac{dx^j\wedge dx^k}{l_s^2}
\label{closedvari}
\eea
gives the open string variables
\be
ds^2=d\vec{r}^2=dx_1^2+dx_2^2+dx_3^3;\;\;\; \theta^{ij}=
\alpha \epsilon^{ijk}x^k
\label{openvari}
\ee
again independent of the value of $l_s$! Thus we can say that $l_s\rightarrow
0$ gives noncommutative geometry with these variables, however we don't need 
to take such a limit to obtain the same open string variables. In particular,
we can take the case of $\alpha^2 r^2/l_s^4\ll 1$ for the whole Universe, 
in which case both the open and the closed string variables have approximately 
flat metric and constant $d\theta$ and $dB$. That means that we choose to 
neglect the difference between the open and closed string metrics (or 
rather, we will analyze the case when this is true). 

The $l_s$ independence of 
the open string variables we take it to mean that $\alpha '$ corrections 
will still respect the noncommutative structure. In other words, if we work 
at energies below the string scale, the $\alpha '$ corrections should decouple
even though we don't have the Seiberg-Witten scaling of the variables. 
This assumption of course deserves further tests, but we will consider it 
plausible at this point and consider that at energies much smaller than 
$1/l_s$ we will have noncommutative geometry. 

But what kind of noncommutativity is the one we found? In \cite{cs} it was 
argued that constant H field means nonassociativity. 

However, one can check that $\theta^{ij}$ in (\ref{openvari}) (as 
well as the one in (\ref{openvar})) satisfies the associativity 
condition found in \cite{cs},
\be
\theta^{il}\partial_l \theta ^{jk} + cyclic[ijk]=0
\label{assoc}
\ee
meaning that the star product defined with it is still associative! 
How can this be? After all, the argument in \cite{cs} is very simple:
the left hand side of the associativity condition (\ref{assoc}) can be 
rewritten as 
\be
\theta^{il}\theta^{jp}\theta^{kq}3\partial_{[l}\theta^{-1}_{pq]}
\ee
and in the Seiberg-Witten limit $\theta^{-1}=B$, and $3dB=H$. Of course,
as \cite{cs} noticed, this is only valid if $\theta$ is invertible, but 
that is not true by definition for a 3 dimensional antisymmetric matrix, 
as we noticed. Moreover, we chose a spherically symmetric gauge for B, 
such that $B_{ij}=a \epsilon_{ijk}x_k$. But we can easily check that 
for any gauge we choose (for instance the one in \cite{lnr}, with 
only $B_{23}$ 
nonzero, if $H_{123}$ is constant, the associativity condition will be 
satisfied!

However, 
$\theta^{il}\partial_l \theta^{ij}$ is still nonzero, meaning that the 
various star products can be different. Indeed, for constant $\theta$, 
one trades the noncommutativity for the usual star product of functions,
\be
f \star g = e^{i\theta^{ij} \partial_i \partial_j'}f(x)g(x')|_{x=x'}
\ee
and if one extends this to the varying $\theta$ case, it matters whether 
we take $\theta (x)$ or $\theta (x')$ in the exponent (or a different 
combination). A well defined prescription for a star product was given 
by Kontsevich \cite{kont} and up to second order in $\theta$ it is 
\bea
f \star g& =& fg + \hbar \theta^{ij} \partial_i f \partial _j g
+ \frac{\hbar ^2}{2} \theta^{ij} \theta^{kl} \partial_i\partial_k f
\partial_j \partial_l g\nonumber\\
&&+ \frac{\hbar^2}{3} (\theta^{ij}\partial_j \theta^{kl})
(\partial_i\partial_k f\partial_l g - \partial_k f\partial _i \partial_l g)
+{\cal O}(\hbar^3)
\eea
and in fact the different between all possible star products is 
given by $\theta^{il}\partial_l \theta^{ij}$ terms. 

Still, when acting on spherically 
symmetric functions $f(r)g(r)$, $\theta^{il}\partial_l \theta^{ij}$
 also gives zero, meaning that the 
star product on spherically symmetric functions is still unique! 

{\bf Open string cosmology}

For completeness, let's further consider  what happens when we go to 
open string metrics.

The metric in (\ref{openv}) can be rewritten as
\be
ds^2= dr^2 + r^2 (1+r^2)d\Omega^2 = (1+r^2) (\frac{dr^2 }{1+r^2}+ r^2 d\Omega
^2)
\ee
that is, conformal to the open FRW model. Indeed, the FRW metric is 
\be
ds^2=-dt^2 + a(t)^2  dl^2 ;\;\;\; dl^2 =\frac{dr^2}{1-kr^2}
+ r^2 d\Omega^2
\ee
where $k=+1$ is closed 3d Universe, $k=0$ is critical (flat), and $k=-1$
is open. Other forms of $dl^2$ for the open Universe are
\be
ds^2 =\frac{d\vec{r}\, '^2}{(1-\vec{r}\, '^2/4)^2}=
\frac{dw_1^2+dw_2^2+dw_3^2}{w_3^2}
\ee
none of which are manifestly homogenous. A space conformal to the homogenous 
space like the one we have is however not homogenous (translationally 
invariant)! The transformation 
of coordinates is $r= r'/(1+k r'^2/4)$ and it gives ($k=-1$)
\be
ds^2= (1+r^2) (\frac{dr^2 }{1+r^2}+ r^2 d\Omega^2)= [\frac{1+r'^2 /4}{
1-r'^2/4}]^2\frac{dr'^2+r'^2 d\Omega^2}{(1-\vec{r}'^2/4)^2}
\ee

One could also analyze the constant spatial H field in 
the general homogenous Universe (FRW) in $r'$ 
variables, i.e. for closed string variables 
\be
ds^2=\frac{d\vec{r}'^2}{(1+k\vec{r}'^2)^2}=\frac{d\vec{r}'^2}{a};\;\;\; 
H_{123}=3
\ee

Then we get the open string variables
\bea
&&ds^2=\frac{1}{a(r)} [dr^2 + a(r)^2 r^2 (1+a(r)^2 r^2)d\Omega^2]\nonumber
\\&&
\theta^{ij} =-\frac{a(r)^2}{1+a(r)^2 r^2}\epsilon^{ijk} x^k
\eea
In the calculation we used  $a(r)=(1+kr^2)^2$ since we could generalize it 
to any a(r). 

So why is the open string variables space not homogenous? In both 
cases, we start with a homogenous (translationally invariant) metric and 
$H_{123}$! But $B_{ij}$ is not homogenous: a translation is equivalent to 
adding a constant B field! In any case, we were interested in having 
minimal variation between the open string and closed string metrics, 
so approximate homogeneity should still hold. But is this automatic?

{\bf Experimental constraints}

Before deciding whether this fact contradicts experimental observations, 
we have to see what are the constraints that observational cosmology 
puts on the H field. 

First, let us see what is the maximal possibility. 
If the energy density of a constant H field is of the order of the 
cosmological constant today 
(which is the maximum allowed value), then $H_{123}^2
\sim \Lambda\sim 10^{-123}M_P^4\sim (10^{-31}M_P)^4$, thus $\sqrt{H_{123}}
\sim 10^{-31}M_P= 10^{-12}GeV=1meV$. 

But then the noncommutativity in (\ref{openv})
(in open string variables) is (with 
$B_{12}= H_{123} x_3$ and dropping the indices, reintroducing dimensions)
\be
[x,x]= \theta= \alpha '\frac{\sqrt{\alpha '} H x}{1+ \alpha ' H^2 x^2}
\ee
and if $\sqrt{\alpha '}\sim l_P$, then $\alpha ' H\sim 10^{-61}$, and
thus on the scale of the Universe, i.e. 
for $x\sim L_0\sim 10^{61}l_P$, we have $[x,x]\sim l_P^2$ (the 
size of the Universe is about
$L_0\sim ct\sim 10^{26}m= 10^{61}l_P$; $l_P\sim 10^{-35}m$). 
Then just on the scale of the Universe we get deviations from homogeneity
(translation invariance), since the factor $\alpha ' H^2 x^2$, that 
quantifies the deviations from homogeneity in both the metric and $\theta$
becomes equal to 1 just at the size of the Universe. 
That might still be important for cosmology, but we will examine it 
in detail in the next section, and the result will be that H is much smaller 
than the maximal value, thus the deviations from homogeneity are negligible, 
and for all intents and purposes one can take $\theta^{ij}=l_0 \epsilon
^{ijk} x^k$ to be the only effect of the constant H on the open string 
variables, with $l_0$ the noncommutativity length scale.

\section{Cosmological relevance of H field and experimental data}

In this section we will study the cosmological implications of adding 
an energy-momentum tensor for the H field ($T_{\mu\nu}(H)$ of the previous 
section) to the one for matter, radiation and cosmological constant. 

We have seen in the previous section that the constant $H_{123}$ (or 
the constant 4 dimensional dual $\tilde{H}_0$) energy momentum tensor
gives equal positive energy and pressure (stiff matter). 

{\bf FRW models}

The general FRW equations are
\bea
&&\frac{\ddot{a}}{a}=-\frac{4\pi G}{3}(\rho + 3 p)\label{exp}\\
&& \frac{\dot{a}^2}{2}- \frac{4\pi G}{3} \rho a^2 = -\frac{k}{2}
\label{accel}
\eea

The two are related by the conservation equation (except for the integration
constant $-k/2$)
\be
\dot{\rho}+ 3\frac{\dot{a}}{a} (\rho + p)=0
\label{cons}
\ee

For k=0 (flat Universe), we have ($p=\gamma \rho$)
\bea
&&a\sim t^n, \;\;{\rm then} \;\;
(\ref{accel}) \Rightarrow \rho = \frac{3n^2}{8\pi G t^2}\nonumber\\
&& (\ref{cons})\Rightarrow n= \frac{2}{3(1+\gamma)}
\eea

If the H field would dominate, we would have $\gamma=1 $ in the previous.
A closed Universe ($k=-1$) with H field dominating is not very plausible 
cosmologically (as it would contradict observation), so let us instead 
examine a closed Universe, with $k=-1$.
For $k=-1$ we have (assuming as usual an equation of state $p= \gamma \rho$)
\bea
&& \frac {\ddot{a}}{a}= -\frac{4\pi G}{3} \rho (1+3\gamma);\;\;\;
(\frac{\dot{a}}{a})^2 -\frac{8\pi G}{3}\rho=\frac{1}{a^2}\nonumber\\
&&\Rightarrow (\frac{\dot{a}}{a})^2 +\frac{\ddot{a}}{a}\frac{2}{1+3\gamma}
=\frac{1}{a^2}
\eea 

As an example of open Universe,  $\gamma = 1/3$ implies $ d(a\dot{a})/dt=1$, 
thus
\be
a(t)= \sqrt{t^2+2Ct+2D};\;\;\; |\rho |=\frac{3}{8\pi G}\frac{2D-C^2}{
(t^2+2Ct+2D)^2}
\ee

By redefining t we can put C=0, and $a_0=a(t=0)=\sqrt{2D}$, thus
\be
a=\sqrt{t^2+a_0^2}\rightarrow t;\;\;\; 
|\rho|=\frac{3}{8\pi G}\frac{a_0^2}{(t^2+a_0^2)^2}\rightarrow \frac{3}{8\pi G}
\frac{a_0^2}{t^4}
\ee

{\bf H field as stiff matter} 

In reality, for constant $H_{123}$ 
we have $\gamma=1$, and thus in an H-field dominated Universe we 
would have $\dot{a}^2+a\ddot{a}/2=1$, which 
seems hard to solve.
But we approximate at late times
\be
a(t)=t+\frac{\alpha}{t^b}+...
\ee
and matching coefficients we get $b=3$, thus
\bea
&& a(t)=t+\frac{\alpha}{t^3}+...\nonumber\\
&& \rho\simeq \frac{9}{4\pi G} \frac{(-\alpha)}{t^6}
\eea
meaning that $\alpha <0$.

At early times, we get 
\bea
&&a\simeq a_0 + a_0 H_0t +\frac{t^2}{a_0}(1-H_0^2 a_0^2)\nonumber\\
&& \rho\simeq \rho_{cr,0}\frac{H_0^2 a_0^2-1}{H_0^2 a_0^2}
\eea

For comparison, let us look at another case of open Universe, 
AdS space ($\Lambda =-|\Lambda|$), which has constant acceleration. 
We have 
\bea
&& \frac{\ddot{a}}{a}=-\frac{|\Lambda|}{3};\;\; \; (\frac{\dot{a}}{a})^2 =
\frac{1}{a^2}-\frac{|\Lambda|}{3}\Rightarrow a\leq \sqrt{\frac{3}{|\Lambda|}}
\nonumber\\
&&a(t)=\sqrt{\frac{3}{|\Lambda|}}\sin \sqrt{\frac{|\Lambda |}{3}}t
\eea
thus in that case we have a maximum size of the Universe. 
For $\rho_H$, at large times we get a rapidly dropping acceleration 
instead (like $1/a^6$):
\be
8\pi G \rho_H=\frac{A}{a^6}>0
;\;\;\; \frac{\ddot{a}}{a}=-\frac{2}{3}\frac{A}{a^6};\;\;\;
(\frac{\dot{a}}{a})^2 \simeq \frac{1}{a^2}+\frac{A}{3}\frac{1}{a^6}
\ee

But let us now understand what happens when the H field is added to other 
types of matter (dust, radiation, cosmological constant). 
From the conservation 
equation it should be clear that we have {\bf for individual components}
\be
\rho_i\sim \frac{1}{a^{3(1+\gamma)}}\Rightarrow
\rho_{\Lambda}\sim 1 ;\;\;\; \rho_R\sim \frac{1}{a^2};\;\;\;
\rho_m\sim \frac{1}{a^3};\;\;\; \rho_{rad}\sim \frac{1}{a^4};\;\;\;
\rho_{stiff}\sim \frac{1}{a^6}
\ee
and thus H behaves like stiff matter, with $\gamma =1$.
These behaviours are independent of what type 
of matter dominates, i.e. of what $a(t)$ actually is. Remember that 
if the component with $\gamma$ dominates, then 
\be
a(t)\sim t^{\frac{2}{3(1+\gamma)}};\;\;\; (\rho\sim \frac{1}{t^2})
\ee
thus for matter domination we have $a(t)\sim t^{2/3}$, for radiation domination
 we have $a(t)\sim t^{1/2}$,
for $\Lambda $ domination we have $a(t)\sim e^{Ht}$, for curvature domination
 we have $a\sim t$ (open Universe at late times).

Observe that
\be
\rho_H\sim H^2\sim \frac{1}{a^6}\Rightarrow H_{123}^2 (g^{-1})^3\sim 
H_{123}^2 a^{-6}\sim \frac{1}{a^6}\Rightarrow H_{123}\sim {\rm const.}
\ee
Thus indeed constant $H_{123}$ is compatible with the equations of motion, 
as it generates stiff matter, that in turn implies $H_{123}$ constant. 

Also note that for a constant H field in a matter dominated flat Universe
(at this moment, we go from matter dominated to acceleration - $\Lambda$?-
dominated cosmology), we 
can solve the FRW equations and find that 
\bea
&& a(t)\simeq a_0[6\pi G \rho_m^0]^{1/3}t^{2/3}(1-\frac{\beta}{t^2})\nonumber\\
&& \Omega _H=\frac{\rho_H(t)}{\rho_{cr}(t)}\simeq \frac{3\beta }{t^2}
 \ll 1\nonumber\\
&& \Rightarrow \rho_H \propto
 \frac{1}{t^2}\times H^2(t)\sim \frac{1}{t^2}\times
\frac{1}{t^2}\sim\frac{1}{t^4}\sim \frac{1}{a^6}
\eea
as expected. 

{\bf Experimental constraints}

Is it possible to have a time-dependent $H_{123}(t)$? 
 The equation of motion 
would be $\partial^{\mu}H_{\mu\nu\rho }=0$ and the Bianchi identity 
$\partial_{[t}H_{123]}=0$. 
Clearly $H_{123}=$constant satisfies both. But if we 
put $H_{123}(t)$ and the only one nonzero, the equation of motion is 
still satisfied (it reduces to $\partial^iH_{ijk}=0$, which is true), but 
the Bianchi identity is not true anymore: it would imply $\partial_t H_{123}=
0$. One could try $H_{tij}\ll H_{123}$ and see if that works, but it 
doesn't. Indeed, that would imply 
\bea
&&\dot{H}_{123}\sim \partial_k H_{tij} \sim \frac{H_{tij }}{L_{Universe}}
\ll \frac{H_{123}}{L_{Universe}}\nonumber\\
&& \Rightarrow \frac{1}{t}= \frac{\dot{H}_{123}}{H_{123}}\ll \frac{1}{
L_{Universe}}
\eea
which is not true!

In conclusion, the only thing that works is that the H field is constant, and 
then it behaves like stiff matter $\rho\sim 1/a^6$, thus if it is to be 
comparable to the cosmological energy density now, it should have been 
dominant before (as it decreases as $1/ a^6$, much faster than matter and 
radiation). That cannot be true, as it would violate all the established 
cosmology. 

Instead, one has to have $\rho_H$ of the order of the overall 
$\rho$ at the Planck scale $M_P$ or the string scale ${\alpha '} ^{-1/2}$
($\rho_H<\rho$ not to disturb usual cosmology and $\rho_H\sim \rho$ from 
naturalness). Then, at the current time, the energy density of the H field 
will be completely negligible, thus will most likely be irrelevant for 
cosmology. 

But that is not entirely excluded from interesting experimental consequences.
It cannot drive the current acceleration, for that we actually need $
\Lambda$, since $\rho_H$ drops quickly, and it cannot be a significant 
contribution to $\rho$, since it would have dominated in the past. But it 
could give noncommutativity at a cosmological scale, as we saw,
without affecting cosmology:

If the string scale is the lowest possible, i.e. if 
$\sqrt{\alpha '}\sim (10 TeV)^{-1}\sim 10^{15} l_P$, in order to have 
 $[x,x]\sim l_P^2$ at the size of the Universe, $L_0\sim 10^{61}
l_P$, one needs 
$H\sim 10^{-106} M_P^2$, thus $\rho_H\sim H^2\sim 10^{-212} M_P^4$ now. 
As it drops as $1/a^6$, it would have been of order $M_P^4$ at $(a_0/a)\sim 
10^{35.3}\sim e^{81}$, i.e. 81 e-foldings ago. More relevant maybe, it would
have been equal to $1/\alpha '^2= 10^{-60}M_P^4$ at $(a_0/a)\sim 
10^{25.3}\sim e^{58}$, i.e. 58 e-foldings ago, thus it could be of just about 
the right order of magnitude to be present, but not have contributed to 
cosmology, as about 50-60 e-foldings are necessary for inflation anyway.

Also, then the only scale relevant for noncommutativity (the length scale 
in $\theta^{ij}= l_0 \epsilon^{ijk}x^k$ is 
\be
l_0\equiv \sqrt{\alpha '} \alpha ' H= 10^{45}\times 10^{-106} l_P
= 10^{-61} l_P\sim l_P^2/L_0
\ee

\section{Noncommutative faster than light soliton solutions}

In \cite{hi} it was shown that in the case of constant noncommutativity 
(coming from constant B field), there exist noncommutative gauge field 
string-like solitons (defined in 2 spatial 
dimensions, with trivial extension in the third), which can have arbitrarily 
high velocities (larger than c), because of the spontaneous breaking of 
Lorentz invariance by the choice of $B_{23}$, say. The arbitrary velocity 
is then in the (2,3) plane, and this seems to provide an explicit breaking 
of rotational invariance (not just Lorentz invariance), which seems hard to 
understand given the experimental observation of perfect rotational invariance
of cosmology. The details of the construction are given in the 
Appendix. We try to parallel that construction here, in  order to find 
faster than light solitons in the rotationally symmetric case, that doesn't 
break further Lorentz invariance othen than the one broken by the FRW 
cosmology.

{\bf Representing the algebra and finding the action}

We want to work on the 3d space with noncommutativity $\epsilon^{ijk} x^k$, 
(we put for the moment the scale $l_0$ to 1) i.e. 
\be\
[X^i, X^j]= i \epsilon^{ijk} X^k
\ee
in other words, on the SU(2) space, or space of quantum angular momenta. 
As is well known, with $X^{\pm}= X^1\pm iX^2$ one can rewrite the algebra as
\be
[X^3, X^{\pm}]= \pm X^{\pm};\;\;\; [X^+, X^-]=2X^3
\ee
and then one can define states $|jm>$ that are eigenvalues of $\vec{X}^2$ 
and $X^3$, which commute. Then $\vec{X}^2|jm>= j(j+1)|jm>$, thus by 
definition $r=\sqrt{j(j+1)}$ can be thought of as the radius in spherical 
coordinates. We see then that we must think of acting on the space of all 
possible representations of SU(2), i.e. arbitrary j. Thus the space is 
like the fuzzy sphere, except the radius (representation of SU(2)) is 
a radial coordinate of space.

It is straightforward to show that the algebra can be represented in terms
of commuting coordinates $x_i'$ and their derivatives as 
\be
X^i= -i\epsilon^{ijk}x'_j \partial '_k
\label{repre}
\ee
for instance 
\be
X^1=-i(x'^2\partial '_3-x'^3\partial '_2)
\ee
Now we would like to define also derivatives on the noncommutative
 space, by putting 
$[\partial_i, X^i]=1$ (no summation). But that is not enough: We need to 
satisfy consistency conditions, derived from applying derivatives to the 
algebra and using the Jacobi identities
\be
[a, [b,c]]=[[a,b],c]+[b, [a,c]]
\ee
Then we obtain, for instance,
\be
[[\partial_1, X^2],X^3]+[X^2, [\partial_1, X^3]]=[\partial_1, [X^2,X^3]]=
i
\ee
We first want to take care of the definition of derivatives, i.e. 
$[\partial_1, X^1]=1$. We  note the action of $x\partial_y-y\partial_x$
on a few combinations:
\be
xy\rightarrow x^2-y^2;\;\;\; x^2-y^2\rightarrow -4xy;\;\;\; 
x^2+y^2\rightarrow 0;\;\;\; x/y\rightarrow -(1+x^2/y^2)
\ee
and find that 
\be
(x\partial_y-y\partial_x) (-arctan (x/y))=1
\ee
and then we can represent the derivative as 
\be
\partial_1= i\; arctan \frac{x'_2}{x'_3}+\alpha x'_1+\beta x'_1\partial_1'
+\gamma X^1 +f((x'_2)^2 +(x'_3)^2)
\label{deriva}
\ee

We can check that this satisfies not only $[\partial_1, X^1]=1$, but also 
the consistency conditions. This leaves however open the value of $[\partial_i,
\partial_j]$, which as we can see in the Appendix is very important for the 
construction of solitonic solutions. We will thus decide on the correct 
representation of derivatives (and thus on $[\partial_i,\partial_j]$) 
when we build solutions. 

Note that at $\alpha =f=0$ both the coordinates $X^i$ and the derivatives 
$\partial_i$ are represented on a unit radius 
commuting 2-sphere (in terms of the phase 
space, i.e. coordinates and derivatives). We observe this by noting that 
scaling $x'_i\rightarrow \lambda x'_i$ doesn't affect $X^i$ and $\partial_i$.

We further observe that in this representation, $X_i$ is actually $-i \partial
/\partial \phi_i$, where for instance 
$\phi_1$ is the angle in the plane formed by the 
commuting cartesian coordinates on $S_2$, $x' _2,x' _3$, etc. Of course, 
there are only two independent angles, thus the three angles are related 
(hence the noncommutation relations). Also, at $\alpha=\beta=\gamma =f=0$, 
the derivatives are just $\partial_i=-i\phi_i$, thus $[\partial_i, X_i]=
-[\phi_i, \partial/\partial \phi_i]=1$. 

The introduction of derivatives $\partial_i$ implies however also the need 
to extend the usual understanding of the $|jm>$ representation. Let's see 
this (in the case $\alpha=\beta=\gamma =f=0$) for the simplest operator, 
$X_3$, and its conjugate $\partial_3$. We have two representations, the 
$|jm>$ representation, and the (``Fourier transformed'') unit radius 
commuting $S_2$ 
representation, in terms of $(\theta, \phi)$ angles. In terms of $|jm>$ 
we have as usual
\be
X_3|jm>=m|jm>
\ee
which translates into the (spherical harmonic) representation of $Y_{jm}
(\theta, \phi)$ as 
\be
X_3 Y_{jm}(\theta, \phi)= -i \frac{\partial}{\partial \phi}Y_{jm}(\theta, \phi)
=mY_{jm}(\theta, \phi)
\ee
and relies upon our abstract expression (\ref{repre}), written as 
\be
<\theta ', \phi '|X_3|\theta, \phi>= -i \frac{\partial}{\partial \phi}
\delta (\theta-\theta ') \delta(\phi- \phi ')
\ee

But now going backwards, we defined on $S_2$ the derivatives as 
\be
<\theta ', \phi '|\partial_3|\theta, \phi>=-i\phi 
\delta (\theta-\theta ') \delta(\phi- \phi ')
\ee
which means  on spherical harmonics
\be
\partial _3 Y_{jm}(\theta, \phi)= -i \phi Y_{jm}(\theta, \phi) 
\ee
which can be obtained from 
\be
\partial_3 |jm>= \frac{\partial}{\partial m} |jm>
\ee
which however is not a usual operator on $|jm>$ states (which would give a 
complex number instead of the function $\partial/\partial_m$), but rather 
is of the type obtained for usual conjugate operators $\hat{X}, \hat{P}$ 
($[i\hat{P}, \hat{X}]=1$) in the x-space basis:
\be
\hat{X}|x>=x|x>;\;\;\; i\hat{P}|x>= \frac{\partial}{\partial x}|x>
\ee

Thus the introduction of derivatives implies an extension of the usual 
understanding of the $|jm>$ basis into a $|x>$-type basis. It becomes 
clear then why we need to specify the derivatives as well in order to define 
the representation space. As mentioned, we will do that when we build 
solutions.

Now we can define covariant derivatives and the YM action for the gauge 
field A in a similar 
manner to the constant noncommutativity case. Define first
\be
C_i=\partial_i +A_i;\;\;\; D_i=[\partial_i+A_i, ]
\ee
and then the field stregth of A is 
\be
F_{ij}= [\partial_i, A_j]-[\partial_j , A_i]+[A_i, A_j]= [C_i, C_j]
-[\partial_i, \partial_j]
\ee

Integration is defined over a sphere by comparing the integral of $1\rightarrow
{\bf 1}=\sum_{jm} |jm><jm|$, thus 
\be
\int_{S}ds_1 ds_2 f(s_1,s_2)= \frac{4\pi j^2}{2j+1}Tr_m f(j,m)
\ee
and thus over the volume of a sphere as 
\be
\int_V d^3 x f(\vec{x})= \sum_{jm}\frac{4\pi j^2}{2j+1} f(j,m)
\ee

Then the action in the $A_0=0$ gauge is (putting also the 
noncommutativity length scale, $[\bar{x}^i, \bar{x}^j]
=il_0\epsilon^{ijk}\bar{x}_k$) 
\bea
&&\frac{2\pi}{g_{YM}^2}\int d\bar{x}_0 d\bar{x}_1 d\bar{x}_2 d\bar{x}_3
F_{\mu\nu}^2=\frac{2(2\pi)^2}{g_{YM}^2l_0}
\int dx_0 \; Tr_{jm}\frac{2j^2}{2j+1}[F_{0i}F^{0i}+F_{ik}F^{ik}]\nonumber\\
&& =\frac{2(2\pi)^2}{g_{YM}^2l_0}
\int dx_0 \; Tr_{jm}\frac{2j^2}{2j+1}
[-\sum_i \partial_t C_i \partial_t C_i+\sum_{i<k}
([C_i, C_k]-[\partial_i, \partial_k])^2]
\label{ncaction}
\eea

Its static equations of motion are
\be
[C_i, [C_i, C_k]]=[C_i, [\partial_i , \partial_k]]
\ee
and the (Gauss law) constraint (coming from varying with respect to $A_0$ 
and then imposing the $A_0=0$ gauge condition) is
\be
[C_i, \partial_t C_i]=0
\ee
(We can think of it roughly as $[D^i, F_{0i}]=0$).

{\bf Building solutions}

We will now try to systematically analyze possible solutions analogous to 
the ones in \cite{hi}. 

If we have an operator S satisfying as in the constant noncommutativity case
(see the Appendix)
\be
S S^+=1;\;\;\; S^+S= 1-{\cal O}
\ee
then we can look for a solution of the type 
\be
C_i= S^+ u_i S+l_i {\cal O}
\ee
where $l_i$ are numbers, that can be identified with the positions of the 
solutions if $u_i$ behaves as $X_i$ (by an argument 
analogous to the one given in the Appendix for the constant noncommutativity 
case),  and the equations of motion then reduce to (if $[{\cal O}, [\partial_i,
\partial_j]]=0$)
\be
S^+[u_i, [u_i, u_j]]S=[S^+u_i S, [\partial_i, \partial_j]]
\label{geneom}
\ee
and we want to look for a solution with nonzero field strength
\be
F_{ij}= [C_i, C_j]-[\partial_i, \partial_j]= S^+[u_i, u_j]S-
[\partial_i, \partial_j]
\ee
such that one has nonzero magnetic flux. For a static solution, the Gauss 
constraint is automatically satisfied.

In a spherically symmetric situation, one would define the magnetic flux 
in the commutative case as
\be
q= \int_{S_2}(ds_1 ds_2)\epsilon ^{ijk} F_{ij} \frac{x^k}{r}
\ee
where $ds_1ds_2$ is the area element on the sphere, and the magnetic flux a
 doesn't depend on the radius of the sphere chosen.
We can check that this formula reproduces the charge of a monopole
($B^k=\epsilon^{ijk}F_{ij}= q x^k/(4\pi r^3)$). 
Before we generalize this to our 
noncommutative case, we have to understand the basis of states, $\{ |jm>\}$. 
The identity operator is 
\be
1=\sum_{j\in N/2, |m|\leq j}|jm><jm|
\ee
where the sum can be thought of as summing over a representation of SU(2)
(a fuzzy sphere, characterized by m) and then over the ``radius'' j (over 
representations of SU(2)). 

A monopole solution in this space would have then 
\be
B^k=\epsilon^{ijk} F_{jk}= \frac{q X^k}{4\pi |X|^3}
\ee
thus, e.g.
\be
B^3= \frac{q X^3}{4\pi |X|^3}=\frac{q}{4\pi}\sum _{jm} \frac{m}{j^3}|jm><jm|
\ee

We can thus represent the magnetic flux by 
\be
\frac{4\pi}{2}(Tr_m(F_{ij} \epsilon^{ijk}X_k))_{j=const.}
\ee
where now we trace over an irrep (j=constant), but the result 
 should be independent 
of j (corresponding to the radius of the sphere), at least at large j. This 
clearly gives the correct result for the monopole above. According to 
our definition of sphere integration, 
there should be also an overall $2j/(2j+1)$ that dissapears 
in the large j limit, but we will neglect it, as we would have also an 
indeterminacy in the definition of $r$, which should really be $\sqrt{\vec{X}
^2}=\sqrt{j(j+1)}$ instead of j, but again the difference vanishes at infinity.

Let us consider what representation of the derivatives $\partial_i$ 
(which we saw have to be of the type (\ref{deriva})) and of $u_i$ we can have. 
If we choose on the unit radius commuting $S^2$
\be
\partial_1= i\; arctan \frac{x'_2}{x'_3}+\alpha X_1,\;\;\; etc.
\ee
we get 
\be
[\partial_1, \partial_2]= i\alpha^2 X_3- \alpha \frac{x_1'x_2'((x_1')^2
-(x_2')^2)}{((x_2')^2+(x_3')^2)((x_1')^2+(x_3')^2)}
\ee
and 
\be
[\partial_1,[\partial_1, \partial_2]]= \alpha^3 X_2+i\alpha^2 x_1'x_3'
[\frac{(x_1')^2+(x_3')^2+2(x_2')^2}{((x_1')^2+(x_3')^2)}
-\frac{1}{x_2'^2 +x_3'^2}]
\ee
whereas if we choose 
\be
\partial_1= i\; arctan \frac{x'_2}{x'_3}+\beta x_1'\partial_1',\;\;\; etc.
\ee
we get 
\be
[\partial_1,[\partial_1, \partial_2]]=i\beta^2 x_1'x_3'\frac{(x_1')^2-(x_3')^2
}{((x_1')^2+(x_3')^2)^2}= 
[\partial_3,[\partial_3, \partial_2]]
\ee
Because the commutator of derivatives is a nontrivial function of $x'$s, 
a solution of the type we wanted, with $u_i=\partial_i$, doesn't work 
for any choice of derivatives, since we will not get an $F_{ij}$ proportional
to $S^+S-1={\cal O}$, as we can easily check from (\ref{geneom}).

However, we can choose the minimal derivative, i.e. 
\be
\partial_1= i\; arctan \frac{x'_2}{x'_3}(=-i\phi_1');\;\;\; etc.
\ee
(thus having $[\partial_i, \partial_j]=0$) and modify just the $u_i$'s. 
We have not found a solution with all $u_i$'s nontrivial, we have to 
put at least one, here $u_3$, to zero:  
\bea
&& u_1= \partial_1 + \alpha_1 X_2\nonumber\\
&& u_2= \partial_2+\alpha_2 X_1\nonumber\\
&& u_3 = 0
\label{us}
\eea
and we see that we want (we will analyze this also later on) a constant 
commutator, thus only one $\alpha$ to be nonzero, e.g. $\alpha_1$
(such that $\alpha_1\alpha_2=0$). However, we 
will keep the discussion general in the following and write $\alpha_i$ 
everywhere assuming that $\alpha_1\alpha_2=0$ and $\alpha_3=0$ and then 
$[u_1,u_2]=-(\alpha_1-\alpha_2)$.
Thus (since also $[\partial_i, \partial_j]=0$), the equations of motion are
easily satisfied. Then in terms of the angles on the unit commuting $S^2$, 
we have $iu_1=\phi_1'+
\alpha_1 \partial/\partial \phi_2'$, $iu_2= \phi_2' +\alpha_2 \partial/\partial
\phi_1'$. 

Moreover, now 
\be
F_{ij}= -(\alpha_i-\alpha_j) S^+S
\ee
is nonzero and then the flux is
\be
\frac{4\pi}{2}(Tr_m(F_{ij}\epsilon^{ijk} X^k)_{j=const}
\ee
but $S^+S$ is diagonal, and $X^1, X^2$ are off diagonal so don't contribute 
to the trace, but $F_{12}$ is the only one nonzero anyway,
hence we get the flux
\be
-4\pi (\alpha_1-\alpha_2)\sum_{m}m(S^+S)_{jm}
\ee
thus it would be constant (independent of the radius j) only if 
$\sum_m m (S^+S)_{jm}$ is constant, but it depends on 
the arbitrary constants $\alpha_i$. 

We will now try to find the operator S. Trying to mimic the constant 
noncommutativity case (see the Appendix), 
it seems natural that the operator S is chosen to be 
\be
S= \sum_{j\in N/2, |m|\leq j}|jm><j+1/2, m+1/2|
\ee
with 
\be
SS^+=1;\;\;\; S^+S= 1-\sum_{j\geq 0}|j+1/2, -(j+1/2)><j+1/2, -(j+1/2)|
\equiv {\bf 1}-{\cal O}
\ee
The full solution would  then be ($C_3=0$)
\be
C_i= S^+ u_i S+l_i \sum_{j\geq 0} |j+1/2, -(j+1/2)><j+1/2, -(j+1/2)|;\;\;\;
i=1,2.
\ee
where $l_i$ are arbitrary real numbers corresponding to the position of 
the flux solution in the 2 angle directions on the fixed radius sphere
directions $\Omega_2$, as on the unit commuting $S^2$ we have
$u_i\sim i\phi_i'$. 
However, it is clear from (\ref{us}) that one really has $u_i \sim (\sum_j
\alpha_j)
\phi_i'$ in some sense, so we should replace 
$l_i\rightarrow (\sum_j \alpha_j) l_i$. 

As usual, we could write down moving solutions by putting $l_i=l_i^0+v_i t$, 
one can easily check that they still satisfy the equations of motion and 
the Gauss constraint (which is nontrivial now), where $v_i $ will be velocities
for motion in the sphere $\Omega_2$. 

Then we would get for the energy 
\be
E= \frac{(2\pi)^2}{g_{YM}^2}\frac{2}{l_0}\sum_j\frac{2j^2}{2j+1}
[ (\sum_j \alpha_j)^2 v_iv_i Tr_m[{\cal O}^2]+ 
(\sum_{ij}(\alpha_i-\alpha_j)^2)Tr_m[({\bf 1}-{\cal O})^2]]
\label{solution}
\ee
Note that since we assumeed $\alpha_i\alpha_j=0$, to have galilean 
invariance we would only need to replace $ ({\bf 1}-{\cal O})^2$ by 
${\cal O}^2$, however the energy is infinite, independent on the $\alpha$'s.

More importantly however, for this solution $\sum_m m(S^+S)_{jm}=j$, thus 
the flux is not constant.

By comparison, for the case in \cite{hi} we could also have a solution with 
(it still solves the equations of motion)
\be
C=S^+\alpha a^+S+\alpha l_0|0><0|
\ee
but then we can check that the flux $Tr(F)$ is not finite, and the only 
finite case is $\alpha=1$. In that case we would have
 $E=|\alpha|^2 m(1+v_i\bar{v}_i)$. 

To get a good solution, we can try to 
relax the requirement that $S^+S=1-{\cal O}$.
The problems of varying flux and/or
non-galilean invariance can be solved simultaneously if $S^+S$ has a single 
representative at fixed j, and equals the term added to C, multiplying
$l_i$, i.e. $S^+S={\cal O}$. 
We can obtain such a situation if we choose for S to map states 
ordered by increasing j and increasing m into states of increasing J, but fixed
M. Specifically, one can choose
\be
S=\sum_{jm} |jm><J= [\frac{1}{2}(\sum_{j'<j}\sum_{m'=-j'}^{j'}1)+\frac{
(j+m)}{2}], M|
= \sum_{jm} |jm><j (j+1/2)+\frac{(j+m)}{2}, M|
\ee
Then, the solution is ($C_3=0$)
\be
C_i=S^+u_iS+\alpha_1
(l_i^0+v_it)\sum_{J}|J,M><J,M|= S^+u_iS+\alpha_1(l_i^0+v_it){\cal O};
\;\;\; i=1,2.
\ee
where $S^+S={\cal O}$, and as we saw from (\ref{us}) and we will confirm 
from the holographic dual in the next section, the velocities are  
on the sphere.
For M, one can choose a constant value. Let us analyze  M=0 and M=1. 
Choosing M=1
(and only $\alpha_1$ nonzero) we get a solution with nonzero flux
\be
F= \frac{4\pi}{2}(Tr_m(F_{ij}\epsilon^{ijk}X_k))_{j=const.}
= 4\pi Tr_m (F_{12}X_3)= -4\pi \alpha_1 
\ee
(and independent of J, as it should!) but it still has infinite energy 
\bea
E&=& \frac{(2\pi)^2}{g_{YM}^2}\frac{2\alpha_1^2}{l_0}\sum_j\frac{2j^2}{2j+1}
[ v_iv_i Tr_m[{\cal O}^2]+ 
Tr_m[{\cal O}^2]]\nonumber\\
&=&\frac{(2\pi)^2}{g_{YM}^2}\frac{2\alpha_1^2}{l_0}
\sum_j \frac{2j^2}{2j+1} (1+\vec{v}^2)
= \frac{(4\pi \alpha_1)^2}{2g_{YM}^2l_0}\sum_j 
\frac{2j^2}{2j+1} (1+\vec{v}^2)
\label{ene}
\eea
 Note that in any case, 
for the static solution, $\alpha_1$ is now arbitrary, allowing for arbitrary 
mass and flux! The flux should naively be automatically quantized
(like in the constant noncommutativity case), but maybe the particular 
noncomutativity precludes quantization of the soliton flux, or maybe if we have
nonzero flux we need to impose by hand $4\pi 
\alpha_1=1$ which would solve the problem. 
Also note that we can think of the solution as a ``tube'' of constant energy 
density= string, situated at $m=M$ (by translating back the sum $\sum_{mj}$
into an integral). 

In any case,
that problem is easily fixed by choosing M=0, giving us zero flux, but the 
same energy. But that means that we have a solution for a flux-less soliton 
of ANY mass (even though there is a characteristic mass scale $1/l_0$), and 
moving at any speed! 

Again, note that in the case of constant noncommutativity, 
quantization of the flux and of 
the energy came about because of the condition of finiteness of the flux and 
energy, but now that is not enough, due to the absence of the 1 in $F_{ij}$,
which came from nonzero $[\partial_i, \partial_j]$. 

{\bf Proposed solution}

In conclusion, we have seen that constant flux and galilean invariance 
forced us to have $S^+S={\cal O}$ instead of ${\bf 1}-{\cal O}$ and then 
we obtained solutions with infinite energy and no quantized flux, but also 
flux-less solutions. Flux conservation means that the solutions with flux 
must extend to infinity (the flux cannot dissappear), and that means summing 
over j up to infinity in the energy formula (\ref{ene}), thus obtaining a 
divergent answer. 

However, for the flux-less solution with M=0, there is no such constraint,
and we could in principle cut off the solution at some $j=j_{max}$, 
and correspondingly at some $J=J_{MAX}$, as well as at a $j_{min}$ and 
$J_{MIN}$, so
let us see if we can do this consistently.

We want  ($C_3=0$)
\be
C_i=S'^+u_iS'+\alpha_1
(l_i^0+v_it)\sum_{J_{MIN}\leq J\leq J_{MAX}}|J,0><J,0|=
 S'^+u_iS'+\alpha_1(l_i^0+v_it){\cal O}'
\ee
where $S'^+S'={\cal O}'\equiv {\cal O}_{J_{MIN}\leq 
J\leq J_{MAX}}$. Here $u_1= \partial_1+\alpha_1 X_2, u_2=\partial_2$
(as before) and S' is 
\bea
S'&=&\sum_{m,j_{min}\leq j\leq j_{max}} 
|jm><J= [\frac{1}{2}(\sum_{j'<j}\sum_{m'=-j'}^{j'}1)+\frac{(j+m)}{2}], 0|
\nonumber\\
&=& \sum_{m,j_{min} \leq j\leq j_{max}} |jm><j (j+1/2)+\frac{(j+m)}{2}, 0|
\eea
thus $S' S'^+={\bf 1}_{j_{min}\leq j\leq j_{max}}\equiv {\bf P}_{j_{min}\leq
j\leq j_{max}}$ (projector onto the subspace of radius $j_{min}\leq j\leq 
j_{max}$). Note however that we don't need to have $J=j(j+1/2)+(j+m)/2$ 
as above (this was necessary if we started at $j_{min}=J_{MIN}=0$ as 
before), which would mean that the physical region in j, $(j_{min},j_{max})$ 
and in J, $(J_{MIN},J_{MAX})$ would be wildly different, 
but we could instead start at $j_{min}=J_{MIN}$ and start mapping $|jm>$ 
states onto increasing J states from there on.  
We can then easily check that this 
is still a solution of the equations of motion and Gauss constraint if 
the derivatives $\partial_i$ commute with the projector ${\bf P}$. We have seen
however that $\partial_3$ acts as $\partial/\partial m$ independent of j
(and by rotations the same would be true for the other X's), thus $\partial_i$
do commute with ${\bf P}_{j_{min}\leq j\leq j_{max}}$. Then 
$C_i$ gives a solution that starts at $J=J_{MIN}$,
extends to $J=J_{MAX}$ and is zero outside, 
and the energy formula is still given by (\ref{ene}) 
just that the summation is from $J_{MIN}$ 
up to $J_{MAX}$. The energy of the solution 
will be, if $J_{MAX}\simeq J_{MIN}$, 
\bea
E&=& \frac{(4\pi \alpha_1)^2}{2g_{YM}^2l_0}
\sum_{J=J_{MIN}}^{J_{MAX}}\frac{2J^2}{2J+1} (1+\vec{v}^2)
\nonumber\\
&\simeq &\frac{(4\pi \alpha_1)^2}{2g_{YM}^2l_0}(1+\vec{v}^2)
J_{MAX}(J_{MAX}-J_{MIN})= \frac{(4\pi \alpha_1)^2}{2g_{YM}^2l_0}
(1+\vec{v}^2)\frac{\Delta J}{J} J_{MAX}^2
\label{disp}
\eea
thus  with an arbitrary mass! However, if we consider the solution with 
unit flux (M=1 before) and ignore flux conservation for a moment, then 
$4\pi \alpha_1=1$ and the mass is finite.

\section{Interpretation and dual gravity theory}

In \cite{hi}, the analysis of the soliton in the noncommutative gauge 
theory was paralleled in a holographic description, with the vortex soliton
(with a trivial extension in the third direction) 
being a D1 string situated at $u\rightarrow \infty$ in the holographic 
dual (see the Appendix). Here we would like to see whether we can do something 
similar. 

In \cite{lnr} it was found that a noncommutative SYM with the open 
string variables
\be
ds^2=  -dt^2 + d\tilde{y}_1^2+d\tilde{y}_2^2 +d\tilde{y}_3^2
+dx^2+d\vec{\sigma}_5^2;\;\;\;\;
\theta^{23}=\tilde{\alpha}\tilde{y}_1;\;\;\; 
e^{\phi}= g_s 
\ee
deformed with a harmonic function $H(x)$, with $x$ being one of the 
transverse coordinates, was obtained from the closed string variables
\bea
&& ds^2 = -dt^2 + d\tilde{y}_1^2 +\frac{d\tilde{y}_2^2 +d\tilde{y}_3^2}{
1+(\tilde{\alpha}\tilde{y}_1/l_s^2)^2}+ dx^2+d\vec{\sigma}_5^2
\nonumber\\
&& B= -\frac{\tilde{\alpha}\tilde{y}_1/l_s^2}{
1+(\tilde{\alpha}\tilde{y}_1/l_s^2)^2}\frac{d\tilde{y}_2\wedge d\tilde
{y}_3}{l_s^2};\;\;\;
e^{\phi}= g_s [1+(\frac{\tilde{\alpha}\tilde{y}_1}{l_s^2})^2]^{-1/2}
\label{closedm}
\eea
again deformed with the harmonic function H(x). Near the core x=0 however, 
the solutions were the ones above, H(x) being necessary in order to obtain 
a solution of the equations of motion. 

The holographic dual to the noncommutative gauge theory was found by 
putting D3 branes in the closed string  background and taking a decoupling 
limit. The D3 branes, with harmonic function (near the x=0 core)
\be
H_1\simeq 1+\frac{4\pi g_s N \alpha '^2}{(\vec{\sigma}^2+x^2)^2}
\ee
was found to be 
\bea
&& ds^2 = H_1^{-1/2}
[-dt^2 + d\tilde{y}_1^2 +\frac{d\tilde{y}_2^2 +d\tilde{y}_3^2}{
1+(\tilde{\alpha}\tilde{y}_1/l_s^2)^2H_1^{-1}}]+ 
H_1^{1/2}[dx^2+d\vec{\sigma}_5^2]
\nonumber\\
&& B= -\frac{\tilde{\alpha}\tilde{y}_1/l_s^2}{
1+(\tilde{\alpha}\tilde{y}_1/l_s^2)^2H_1^{-1}}
\frac{d\tilde{y}_2\wedge d\tilde{y}_3}{H_1 l_s^2};\;\;\;
e^{\phi}= g_s [1+(\frac{\tilde{\alpha}\tilde{y}_1}{l_s^2})^2H_1^{-1}]^{-1/2}
\label{dbr}
\eea
and the decoupling limit was $\alpha ' \rightarrow 0$, keeping 
$\tilde{y}_i,\tilde{\alpha}, U=|\vec{\sigma}|/\alpha '$, $X=x/\alpha '$, $g_s
N=\lambda$ fixed. In this limit, $l_s^4 H_1=4\pi \lambda /U^4$= fixed
and then the holographic dual is just the usual one in (\ref{ncdual}), with 
noncommutativity $\theta= \tilde{\alpha}\tilde{y}_1$. 

We have seen that the noncommutative geometry discussed in this paper, 
with open string variables in (\ref{openvari}) is obtained from the closed 
string variables in (\ref{closedvari}), which is the rotational invariant 
form of (\ref{closedm}). As we mentioned, to make (\ref{closedm}) a 
solution at arbitrary x, we need to introduce a harmonic function $H(x)$, 
and most likely the same should be true for (\ref{closedvari}), i.e. the 
solution will be valid only for some of the transverse coordinates $(x,
\vec{\sigma}_5)$ being small, which we will generically call $\vec{x}$. 
We will however assume that this is a solution 
for at least one of the transverse coordinates being arbitrary, in general
called $\vec{\sigma}$. 

Then in order to have a dual to (\ref{openvari}) we need to put D3 branes 
in (\ref{closedvari}). Applying to (\ref{dbr}) the same rotation that applied 
to (\ref{closedm}) gives (\ref{closedvari}), we get 
\bea
&& ds^2 = H_1^{-1/2}[-dt^2+dr^2 +\frac{r^2 }{1+ H_1^{-1} (\tilde{\alpha}r/l_s
^2)^2}d\Omega_2^2] + H_1^{1/2}(d\vec{\sigma}^2+d\vec{x}^2)\nonumber\\
&& B= -\frac{\tilde{\alpha}/l_s^4}{1+ H_1^{-1} (\tilde{\alpha}r/l_s
^2)^2}\epsilon_{ijk} \frac{\tilde{y}^i d\tilde{y}^j\wedge d\tilde{y}^k}{H_1}
\nonumber\\
&& e^{\phi}= g_s [1+ H_1^{-1} (\tilde{\alpha}r/l_s^2)^2]^{-1/2}
\eea
and the harmonic function is 
\be
H_1= 1+\frac{4\pi g_s N \alpha '^2}{(\vec{\sigma}^2+\vec{x}^2)^2}
\ee

The decoupling limit is the same as the one in \cite{lnr}, and one 
finds $l_s^4 H_1= 4\pi \lambda /U^4$, and then 
\bea
&&\frac{ds^2}{l_s^2}= \frac{U^2}{\sqrt{4\pi \lambda}}[
-dt^2 +dr^2 +\frac{r^2}{1+\tilde{\alpha}^2 r^2 U^4/(4\pi \lambda)}d\Omega_2^2]
+\frac{\sqrt{4\pi \lambda}}{U^2}(dU^2+d\vec{X}^2)+\sqrt{4\pi\lambda}d\Omega^2
\nonumber\\
&& B= -\frac{\tilde{\alpha}U^4}{4\pi \lambda} \frac{1}{
1+\tilde{\alpha}^2 r^2 U^4/(4\pi \lambda)}
\epsilon_{ijk} \tilde{y}^i d\tilde{y}^j\wedge d\tilde{y}^k;\;\;\;\; 
e^{\phi}=g_s [1+\tilde{\alpha}^2 r^2 U^4/(4\pi \lambda)]^{-1/2}
\nonumber\\
&& A_{0r}= \frac{2\pi\alpha '}{g_{YM}^2}\frac{\tilde{\alpha}r U^4}{\lambda}
\eea
Note that here we have put also $A_{0r}$. In \cite{lnr} only the NS-NS 
fields were written explicitly, but one can easily check that $A_{\mu\nu}$ is 
there also (the solution was found by performing transformations on a 
D1 solution, carrying $A_{01}$ in that case).  

In the case in \cite{hi}, the dual to the magnetic vortices were D1 strings, 
as they carried gauge flux. In our case however, the solitons are just 
string-like objects living in 
the $(t,r,\Omega_2)$ directions, but they do not carry flux so they are 
not D-branes. We will however calculate first what happens for D1 branes, 
which have fixed D1 tension, and then make the tension arbitrary. The 
action is 
\be
S_{D_1}=
-\frac{1}{2 \pi \alpha '}\int dx^0 dx^1 e^{-\phi} \sqrt{det (g_{\mu\nu}
\partial_a X^{\mu}\partial_b X^{\nu})}+\frac{1}{2\pi \alpha ' }
\int A_{\mu\nu} \partial_0 X^{\mu}\partial_1 X^{\nu} dx^0\wedge dx^1
\label{Daction}
\ee

Let us look at a string oriented along t and r.
 Then the static potential of a string of length L in $\sigma$
will be (S=$-\int dt V(<r>,U)$ and $<r>$ is the average position of the 
string in the r direction)
\be
V(<r>,U)= \int dr \frac{\tilde{\alpha} r U^4}{4\pi \lambda(2\pi g_s)}
[\sqrt{1+\frac{4\pi \lambda}{\tilde{\alpha}^2 r^2 U^4}}-1]
\ee
and the potential becomes constant at $U\rightarrow \infty$, namely 
\be
V(<r>,\infty ) = \frac{1}{2g_{YM}^2\tilde{\alpha}}\int \frac{dr}{r}
\ee

As a result (since V'(U)=0 at infinity), 
as for the solitons in \cite{hi}, these static strings living 
at $U=\infty$ will be solutions of the equations of motion. Notice also 
that then $V'(<r>)=V(<r>)=0$ at $U=\infty$, so the strings will settle 
(and become solutions of the equations of motion) at large $<r>$. 

The complete velocity-dependent action for these strings moving on the 
sphere $\Omega_2$  (since as we saw our solution moves on $\Omega_2$ also)
is found easily to be 
\be
S= -\int dt \int dr \frac{\tilde{\alpha} r U^4}{4\pi \lambda(2\pi g_s)}
[\sqrt{1+\frac{4\pi \lambda}{\tilde{\alpha}^2 r^2 U^4}}\sqrt{1-\frac{
\vec{v}_{\Omega}^2r^2(4\pi \lambda)}{4\pi \lambda+\tilde{\alpha}^2r^2 U^4}}
-1]
\ee
and we see that in the limit $U\rightarrow \infty$ this dispersion relation 
is galilean, i.e. there is no bound on velocity, exactly as in the case 
in \cite{hi}. More precisely, the velocity is bounded by $
4\pi \lambda+\tilde{\alpha}^2r^2 U^4$ which becomes infinite as 
$U\rightarrow \infty$. The energy becomes in the limit
\be
E= \frac{1}{2g_{YM}^2\tilde{\alpha}}\int \frac{dr}{r}(1+\vec{v}_{\Omega}^2r^2)
\simeq \frac{1}{2g_{YM}^2\tilde{\alpha}}\frac{L}{<r>}(1+\vec{v}_{\Omega}^2
<r>^2)
\label{disper}
\ee
where L is the length of the string $\Delta r$ and $\vec{v}_{\Omega}r=\vec{v}$
is the velocity in the sphere directions. We note that this is exactly the 
formula in (\ref{disp}) for the unit flux case ($M=1,4\pi \alpha_1=1$), 
except for a factor of 
$J_{MAX}^2$. We can take it to mean that flux conservation in the 
spherically symmetric case means that 
$J_{MAX}^2$ D1 strings are required (being mostly paired as positive/negative
flux pairs, with a single effective unit of flux) in this noncommutative case.
It is also possible that perhaps due to the peculiar definition of the 
derivatives, in (\ref{ncaction}) $F_{ik}F^{ik}=(1/j^2)([C_i,C_k]-[\partial_i,
\partial_k])^2$ which would also solve the discrepancy, but it is hard to 
see why that would be so. 

For the flux-less solution, the overall tension in (\ref{Daction}) is 
arbitrary, and then there is an arbitrary constant multiplying (\ref{disper}),
giving solutions with arbitrary mass, as obtained in (\ref{disp}).
  
So we have found that the noncommutative gauge theory studied has flux-less 
solutions of arbitrary mass and with galilean dispersion relations, thus 
with speeds arbitrarly higher than the speed of light.

The same comments as in \cite{hi} apply: the velocity in the gauge theory 
is velocity in the open string metric. In the holographic dual, this 
velocity is just coordinate velocity, the velocity measured by a local 
observer is still bounded by the speed of light. 

Causality is not violated because we have a spontaneous violation of  
Lorentz invariance, brought about by the constant $H_{123}$, which defines 
a preferred time slicing, exactly like the FRW metric ansatz (thus phsyically
by the CMBR background). The oddity is of course that the Lorentz violation 
is arbitrarily large (independent on the value of $H_{123}$, as it was 
independent on the value of $\theta$ in \cite{hi}).

\section{Conclusions}

In this paper we have analyzed the consequences of a constant NS-NS H field 
with 4d spatial indices, i.e. $H_{123}$. Such a field has an equal 
positive energy density and pressure, $\rho=p>0$, i.e. stiff matter. 
Cosmological constraints 
on such a tensor are enough to guarantee that open strings and closed 
strings see approximately the same metric over the whole Universe, and the 
noncommutativity is $\theta^{ij} =\alpha \epsilon^{ijk}x^k$. 
We have argued that since for a certain ansatz (\ref{closedvari}) giving the 
noncommutative space (\ref{openvari}) the Seiberg-Witten limit is 
unnecessary, presumably $\alpha '$ corrections should decouple at energies
below the string scale, even though we are not in the Seiberg-Witten limit. 
Even though the H field is constant, the star product defined by $\theta^{ij}$
is associative. Having a constant H field at the Planck scale, in the 
presence of Dp branes needed for the existence of open strings will be 
hard in general, but in this paper we did not deal with the possibility 
of having consistent string models for the Planck era, but seeing what are 
the possible implications of a general model giving a constant H field.  

The H field has positive energy density, thus if it would dominate, the cases
of interest would be flat or 
open Universe, and we showed what happens for an open Universe. 
If it is added to other types of matter, it 
gives a behaviour like stiff matter, i.e. $\rho_H\sim 1/a^6$. That means 
that at this moment, the energy density of H would be negligible, and it 
should have been of the order of the total energy density at an early time, 
like the Planck scale or the string scale. For instance, if $[x,x]\sim l_P^2$
at the scale of the Universe now, and $l_s\sim (10 TeV)^{-1}$, then $\rho
\sim l_s^{-4}$ about 58 e-foldings ago.

The noncommutative space with $\theta^{ij} =\alpha \epsilon^{ijk}x^k$
is a collection of fuzzy spheres, with arbitrary radius, i.e. the radial 
slices are fuzzy spheres. That means that the representation space is 
given by aribtrary $\{ |jm> \}_{jm} $. But one needs to define derivatives 
on this space in order to define gauge theory, and then one needs to 
define also objects like 
$\partial/ \partial m |jm>$, in analogy with $\partial/ \partial x
|x>$ for the usual (commuting) representation space. 
On this noncommutative space we built solutions analogous to the ones in 
\cite{hi}, that can be interpreted as strings extending in a radial direction.
There was no automatic flux quantization condition however, and moreover, 
we found flux-less solutions too. The flux-less solutions thus have 
arbitrary tension (as well as length), and have a galilean dispersion 
relation, thus being able to travel at speeds arbitrarily higher than the 
speed of light. 

In the dual description, the strings are solitonic strings of finite length 
living at $U=\infty$, and their galilean dispersion relation comes from 
the fact that $g_{\Omega\Omega}/g_{00}\rightarrow 0$ as $U\rightarrow 
\infty$, thus because their coordinate velocity is unbounded at $U\rightarrow 
\infty$, even though of course the local velocity is always bounded by c.

The fact that we have solitons of arbitrarily large speeds is as puzzling 
here as in \cite{hi}, the mechanism by which this Lorentz violation 
doesn't contradict causality is somewhat obscured, although of course 
its source is having a spontaneous Lorentz violation by $H_{123}$, which 
makes the usual arguments about choosing moving reference frames and 
creating an a-causal loop invalid, as explained in \cite{hi}. 
Here we have the additional puzzle that the masses of the solitons are 
arbitrary, even though there is a natural energy scale in the theory, 
the noncommutativity scale $1/l_0$. 

Of course the solitons that we have studied in this paper could in principle
be quantum mechanically unstable, at least the flux-less ones, for which 
there is no topological constraint. The same argument used in \cite{hi}, 
that it takes an infinite amount of time to get to $U=\infty$ in the 
holographic dual could maybe be used to argue the fact that they are 
long lived though. 

There are many things to be further studied. The possibility that $H_{123}$ 
actually creates the initial conditions for the start of FRW cosmology 
(as opposed to just being consistent with them) needs to be explored. 
One needs to study whether or not the assumed decoupling of $\alpha '$ 
corrections in this noncommutative case is actually valid. Both the 
noncommutative gauge theory and the solitons we have found deserve 
further study too.

{\bf Acknowledgements} 
I would like to thank Antal Jevicki for useful discussions, Justin 
Khoury for reading the manuscript and spotting a mistake in the first draft, 
and Aki 
Hashimoto for reading the manuscript and having many useful comments.
This research was  supported in part by DOE
grant DE-FE0291ER40688-Task A.

\newpage

{\Large\bf{Appendix A. Constant noncommutativity faster than light soliton.}}

\renewcommand{\theequation}{A.\arabic{equation}}
\setcounter{equation}{0}

\vspace{1cm}

In this appendix we present the details of the faster than light soliton
solution of \cite{hi}. It deals with the constant noncommutativity, 
described in complex coordinates 
\be
[x^2, x^3]= i\theta; \;\;\; 
z=\frac{x^2+ix^3}{\sqrt{2}}\Rightarrow [z,\bar{z}]=\theta
\ee
By rescaling, the algebra of coordinates becomes that of creation/annihilation
operators:
\be
[a, a^+]=1
\ee
Then one can define derivative with respect to the complex coordinates 
as 
\be
\partial \rightarrow -[a^+,\; ];\;\;\; \bar{\partial}=[a, \; ]
\ee
They are uniquely defined and obey 
\be
[\partial , z]=1;\;\;\; [\partial , \bar{\partial}]=1
\ee
where the first condition is imposed as a definition of derivatives, and 
then the second follows as a consequence.

On a scalar field, the covariant derivative will be 
\be
D\phi= \partial \phi +i [A, \phi]=-[C, \phi];\;\;\; 
C= a^+ -iA
\ee
Then one defines the field strength of the gauge field A in the usual way  
\be
F=-[a^+, \bar{A}]-[a, A]+[A, \bar{A}]= [C, \bar{C}]+1
\ee
where the 1 comes from $[\partial, \bar{\partial}]$. 
Thus unlike the commutative case, the field strength is not just the 
commutator of covariant derivatives (or rather, we have to define what that 
means). The YM action (for the gauge field A) on the noncommutative
 space, in the $A_0=0$ gauge  is then
\be
\frac{2\pi}{g_{YM}^2}\int dx_0 dx_1 dx_2 dx_3 F_{\mu\nu}^2=
\frac{2\pi }{g_{YM}^2}\int dx_0 (\int dx_1)\; \theta Tr[
- \partial_tC\partial_t\bar{C}+([C, \bar{C}]+1)^2]
\ee
and the equations of motion and constraints following from it are 
\be
\partial_t^2 C=[C, [C, \bar{C}]];\;\;\;
[C, \partial_t\bar{C}]+[\bar{C}, \partial_t C]=0
\ee
We assume that the $x_1$ direction is trivial, so $\int dx_1=L$ is an 
arbitrary constant (the extension in the $x_1$ direction of the solution,
that will then look like a string soliton in 3+1 dimensions).
The constraints are Gauss law constraints, obtained by varying with respect 
to $A_0$ and then imposing the $A_0=0$ gauge condition (one can think 
of it roughly as $[D_i, F_{0i}]=0$).
 
In the static case they reduce to just 
\be
[C, [C, \bar{C}]]=0
\label{ctnceom}
\ee
States on the noncommutative plane are functions of $z, \bar{z}$, which are 
however represented by creation and annihilation operators. Thus one can 
represent any state as a linear combination of the basis states 
 for the $a, a^+$ 
oscillators, namely $\{ a_n\}_{n\in Z}$. Then the identity on the 
noncommutative plane is represented as
\be
{\bf 1}=\sum_{n \geq 0}|n><n|
\ee
Defining the shift operator 
\be
S= \sum_{i=0}^{\infty} |i><i+1|;\;\;\; S^+= \sum_{i=0}^{\infty}
|i+1><i|
\ee
we see that it satisfies
\be
S S^+={\bf 1};\;\;\; S^+S= {\bf 1}-|0><0|
\ee
Then one finds that a solution of the equations of motion (\ref{ctnceom})
with unit magnetic flux is given by 
\be
C= S^+ a^+ S+\frac{1}{\theta} l_0 |0><0|
\ee
where $l_0$ is an arbitrary complex number, corresponding to the position 
of the flux on the noncommutative plane. Then for this solution one has
\be
\theta F = [C, \bar{C}]+1=|0><0|
\ee

Note that $C=-\partial -i A$, and thus applying a translation to the 
$l_0=0$ solution $C=S^+a^+S=S^+\bar{z}S=-S^+\partial S$
would mean acting with $[exp(l_0 \bar{\partial}),]$ (``Taylor expansion'')
on A, thus acting on C as 
\be
[exp(l_0 \bar{\partial}),C]+[exp(l_0 \bar{\partial}), \partial]=
[exp(l_0 \bar{\partial}),C]- l_0 {\bf 1}
\ee
where we see that by definition $l_0$ corresponds to $\bar{z}$ translation.
The explicit translation acting on $C$ (the first term) is more complicated, 
and probably involves a change of gauge,
but if we think of it as just translating $S^+\bar{z}S\rightarrow 
S^+(\bar{z}+l_0)S$ with $l_0$ a commutative quantity (number), we get 
indeed 
\be
S^+a^+S\rightarrow S^+a^+S-l_0 ({\bf 1}-S^+S)= S^+a^+S-l_0|0><0|
\ee

The magnetic flux through the noncommutative plane is defined as  
\be
\int F_{23} dx^2 \wedge dx^3
\ee
represented now (on the noncommutative plane in oscillator states) by 
\be
Tr_n \theta F
\ee
and we see that the solution has indeed flux equal to 1.

A multi-flux (M units) solution is 
\be
C= (S^+)^M a^+ S^M +\frac{1}{\theta} \sum _{i=0}^{M-1}l_i |i><i|
\ee
with the arbitrary complex numbers $l_i$ representing the flux positions. 
Then, as 
\be
S^M (S^+)^M=1;\;\;\; (S^+)^M S^M=1-\sum_{i=0}^{M-1} |i><i|
\ee
we get 
\be
\theta F= \sum_{i=0}^{m-1} |i><i|\Rightarrow {\rm flux}= M
\ee

The energy of this soliton is 
\be
E= \frac{2\pi \theta }{2 g_{YM}^2}L Tr F_0^2=\frac{\pi L M}{g_{YM}^2\theta}
\ee
for M units of flux. 

For a  moving solution one 
just replaces $l_i^0$ with $l_i=l_i^0+v_it$ and obtains, from the term 
$\partial_t C\partial_t\bar{C}$ in the energy (which changes sign when 
going from S to E!)
\be
E= m(1+v_i \bar{v}_i)
;\;\;\; m= \frac{\pi L M}{g_{YM}^2\theta}
\ee
that is, a nonrelativistic solution energy! Note that here, v is a complex 
number, meaning it is a velocity in the complex (2,3) plane. The velocity 
in the 1 direction is still bounded by c. 

Note that the dispersion relation
appeared because the $l_i$ term has also M terms, thus $Tr [
\partial_t C\partial_t\bar{C}]= Mv_i \bar{v}_i$, same as $Tr [F_0^2]=M$
(M units in both traces).

In the supergravity dual theory to the noncommutative field theory, with 
relevant fields \cite{hitwo,mr}
\bea
&&\frac{ds^2}{\alpha '}=\frac{U^2}{\sqrt{\lambda}}(-dx_0^2+dx_1^2)
+\frac{\sqrt{\lambda}U^2}{\lambda+\Delta^4 U^4}(dx_2^2+dx_3^2)
+\frac{\sqrt{\lambda}}{U^2}dU^2+\sqrt{\lambda}d\Omega_5^2
\nonumber\\&&
e^{\phi}=\frac{g_{YM}^2}{2\pi}\sqrt{\frac{\lambda}{\lambda+\Delta^4U^4}};
\;\; B_{23}= -\frac{\alpha ' \Delta ^2 U^4}{\lambda +\Delta^4 U^4};\;\;
A_{01}=\frac{2\pi}{g_{YM}^2}\frac{\alpha ' \Delta^2 U^4}{\lambda} 
\label{ncdual}
\eea
and $\theta=2\pi \Delta^2$, the dual to the magnetic vortex string soliton 
(the noncommutative soliton above with an arbitrary length in the trivial 
3rd spatial direction)
is a D-string moving in this background. The potential for a D string to move 
up to $U=\infty$ is found to be 
finite, and matches the energy of the soliton in the field 
theory, thus the soliton is actually a D string that reached $U=\infty$. In 
fact, one can check that the limit $U_0\rightarrow \infty$ (the position of the
soliton goes to infinity) takes the role of decoupling limit. 

Then the dependence of the D string action on velocity is found to be 
($x_0\equiv t$)
\be
S\propto\sqrt{\frac{ds^2}{dt^2}}=\sqrt{1-\frac{v^2}{1+\Delta^4 U_0^4/\lambda}}
\ee
where $v^2=(\partial_0 x^2)^2+(\partial_0 x^3)^2$. The complete action is 
actually 
\bea
S &=& -\frac{1}{2\pi \alpha '}\int dx_0 dx_1
e^{-\phi}\sqrt{g_{11}\frac{ds^2}{dt^2}}
+\frac{1}{2\pi \alpha '}\int A_{01} dx_0\wedge dx_1\nonumber\\
&=&-\int dt 
\frac{L\Delta^2 U_0^4}{\lambda g_{YM}^2}(\sqrt{1+\frac{\lambda}{
\Delta^4U_0^4}}\sqrt{1-\frac{v^2}{1+\Delta^4 U_0^4/\lambda}}-1)
\nonumber\\&&
\Rightarrow E(U)\rightarrow \frac{L}{2\Delta^2g_{YM}^2}(1+v^2)
\eea
One can check that for $U_0\rightarrow \infty$ we get the same nonrelativistic
dispersion relation as derived in the noncommutative theory, thus no bound on
 v anymore. The reason we take $U_0\rightarrow \infty$ is that the 
potential at infinity is constant, thus $V'(U)=0$, thus a static string 
at infinity solves the equations of motion.
Thus the fact that the bound on v dissappears is due to the fact that 
$g_{22}/g_{00}=g_{33}/g_{00}\rightarrow 0$ as $U \rightarrow \infty$
in the gravity dual. But one should note that this is just the coordinate 
velocity in the gravity dual, the velocity measured by a local observed in 
the background is still bounded by the speed of light. 

We should also stress that the soliton velocity we refer to is the velocity 
in the open string metric, but the D branes live in the closed string 
metric and they have a relativistic dispersion relation, only in open
string variables we have a nonrelativistic relation.

\end{document}